\documentclass{optica-article}

\journal{opticajournal} 

\articletype{Research Article}

\usepackage{lineno}

\usepackage{graphicx}
\usepackage{amsmath}
\usepackage{booktabs}
\usepackage{tabularx}
\usepackage{mathtools}
\usepackage[algo2e]{algorithm2e} 
\usepackage{multirow}
\usepackage{pifont}
\usepackage{booktabs}
\usepackage{bbding}
\usepackage{multirow}
\usepackage{pifont}
\usepackage{wasysym}
\usepackage{algorithm}
\usepackage{algpseudocode}
\usepackage{arydshln}
\usepackage{float}

\newcommand{\xmark}{\ding{55}}%
\newcommand{\cmark}{\ding{51}}%

\begin{document}

\title{From Hours to Seconds: Towards 100x Faster Quantitative Phase Imaging via Differentiable Microscopy}

\author{Udith Haputhanthri, \authormark{1,2} 
Kithmini Herath, \authormark{1,2}
Ramith Hettiarachchi, \authormark{1,2} 
Hasindu Kariyawasam, \authormark{1,2}
Azeem Ahmad, \authormark{3}
Balpreet S. Ahluwalia, \authormark{3}
Ganesh Acharya, \authormark{4}
Chamira U. S. Edussooriya, \authormark{2,*} and
Dushan N. Wadduwage \authormark{1,*}}

\address{\authormark{1}Center for Advanced Imaging, Faculty of Arts and Sciences,  Harvard University, Cambridge, MA 02138, USA\\
\authormark{2}Department of Electronic and Telecommunication Engineering, University of Moratuwa, Sri Lanka\\
\authormark{3}Department of Physics and Technology, UiT The Arctic University of Norway, Tromsø 9037, Norway\\
\authormark{4}Division of Obstetrics and Gynecology, Department of Clinical Science, Intervention and Technology, Karolinska Institute, Stockholm, Sweden}

\email{\authormark{*}wadduwage@fas.harvard.edu; chamira@uom.lk} 

\begin{abstract*} 
With applications ranging from metabolomics to histopathology, quantitative phase microscopy (QPM) is a powerful label-free imaging modality. Despite significant advances in fast multiplexed imaging sensors and deep-learning-based inverse solvers, the throughput of QPM is currently limited by the speed of electronic hardware. Complementarily, to improve throughput further, here we propose to acquire images in a compressed form such that more information can be transferred beyond the existing electronic hardware bottleneck. To this end, we present a learnable optical compression-decompression framework that learns content-specific features. The proposed differentiable quantitative phase microscopy ($\partial \mu$) first uses learnable optical feature extractors as image compressors. The intensity representation produced by these networks is then captured by the imaging sensor. Finally, a reconstruction network running on electronic hardware decompresses the QPM images. In numerical experiments, the proposed system achieves compression of $\times$ 64 while maintaining the SSIM of $\sim 0.90$ and PSNR of $\sim 30$ dB on cells. The results demonstrated by our experiments open up a new pathway for achieving end-to-end optimized (i.e., optics and electronic) compact QPM systems that may provide unprecedented throughput improvements.

\end{abstract*}

\section{Introduction}

Among the label-free imaging modalities, quantitative phase microscopy (QPM) is a simple but powerful approach, providing important biophysical information by quantifying optical phase differences \cite{3popescu2006,4park2006}. From the phase map, one can further yield quantitative information about the morphology and dynamics of the examined specimens \cite{5fang-yen2007,6amin2007}. In addition to morphology, the measured phase maps can be converted to dry mass of the cells with accuracy that is of the order of femtograms per square microns \cite{7sung2014,8choi2007}. QPM has found many important applications in biomedicine \cite{Park2018} including pathogen screening \cite{10jo2017}, cancer cell classification \cite{Roitshtain2017}, and label-free analysis of histopathology specimens \cite{12majeed2019,Rivenson2019PhaseStain:Learning}. Importantly, label-free histopathology based on QPM has been shown to capture subtle, nanoscale morphological properties of tissues that could lead to early detection of cancer \cite{14Wang2011}. This technique also preserves the tissue sample for further molecular-specific pathological analysis \cite{15Cree2014}. Moreover, recently quantitative phase imaging has even been extended to image the structures of thick biological systems such as zebrafish larval \cite{16Kandel2019}. QPM has also been demonstrated for stain-free quantification of chromosomal dry mass in living cells \cite{17sung2012}, quantification of different growth phases of chondrocytes \cite{Sung2013SizeMicroscopy}, identification of biophysical markers of sickle cell drug responses \cite{19Hosseini2016} and quantification of nuclear mechanical properties that may play roles in cancer metastasis \cite{20Singh2019}.

The first phase imaging mechanism was introduced by Zernike in his phase contrast microscopy \cite{Zernike1942}. Here the phase shifts due to the refractive indices and depth differences in the specimen are converted into detectable intensity variations. Zernike's original design consisted of a phase filter which directly displays phase information by interfering scattered portion of light from an image, with its unscattered portion. Even though the work improved with several extensions \cite{Gluckstad1996, diffphasecontrast}, due to the non-linear dependency between phase and intensity, direct phase contrast techniques are incapable of quantitative phase measurements. QPM techniques overcome this problem by computational inverse reconstruction \cite{Park2018}. A typical quantitative phase microscope consists of an optical system (forward model) and a computational phase retrieval algorithm (inverse model) \cite{Jo2018}. The forward optical system converts undetectable phase information into detectable fringe patterns; from the fringe patterns, the inverse reconstruction algorithm retrieves phase and intensity maps of the specimen. Recent developments in QPM have mostly been focused on improving the inverse reconstruction 
using GPU acceleration \cite{Kim2013,Lim2015, Sung2009}, deep-learning-based inverse solvers \cite{Nguyen2018, Di2020, Zhu2021, Wang2020_y4net, Wang2018_eholonet, Wang2019_ynet}, and illumination patterns optimization \cite{Kellman2019Physics-BasedImaging,Matlock:19}.

Orthogonal to the current developments, the main bottleneck of QPM-based imaging cytometry is the image acquisition speed, which is fundamentally governed by the pixel rate of image sensors. Currently, the pixel rate of a state-of-the-art camera is around $1\times 10^{10}$ pixels/sec. However, the pixel throughput of the front-end optics is virtually unlimited. An image passes through optics at the speed of light and has been the rationale for developing optical signal processing technologies \cite{21yeh1995}. Here we propose to exploit this property to optically compress an image in order to measure the compressed form of the image using a high-speed light detector (such as a high-speed camera). Thus the pixel throughput of the original image would be increased at a rate proportional to the degree of compression. Compressive imaging of biological specimens has been demonstrated before, using random sampling of the linearly projected image space \cite{23Vincent2012}. Better compression, however, may be achieved through learning dataset-specific features of images. To this end, here we propose differentiable microscopy ($\partial \mu$) to identify important image features for compression, learned through machine learning approaches\cite{2022udith}. First, we modeled the compression and de-compression as a single auto-encoder neural network. Afterward, with a large set of target images, we trained the neural network to find a low-dimensional compressed representation of the images. Once trained, the decoder part of the network acts as a decompression algorithm. We then model the encoder part of the network as a learnable optical system to be used as an optical phase feature extractor. More specifically, the optical phase feature extractor encodes phase information at the compressed latent intensity field. The intensity field is then acquired by a photodetector array. The detected intensity feature map is decoded as the phase image by an electronic phase reconstruction network.

Our optical phase feature extractor is motivated by the previous work on all-optically retrieving the phase information through learnable optics \cite{herath2022}. This work modeled phase retrieval as an optimization problem where the optical network's physical parameters were optimized by minimizing the pixel distance between the input phase and the output intensity.  The main limitation of this previous work is the lack of \textit{non-linearity} of the all-optical method. In contrast, here we treat the all-optical linear network as \textit{a feature extractor} which can extract features from both phase and amplitude of the input field. The optical model only has to learn a faithful representation that contains sufficient information to reconstruct the original phase. Combining this all-optical phase feature extractor with the non-linear phase reconstruction network, we realize an end-to-end-learnable non-linear function for the phase retrieval task. In the following sections, we first introduce the proposed end-to-end differentiable compressive QPM (section \ref{sec:results_intro}). Second, we assess the feasibility of phase retrieval using linear compression and non-linear decompression functions (section \ref{sec:results_pre}). Third, we demonstrate \textit{Compressed QPM} using an optical encoder and an electronic decoder. We use learnable Fourier filters (LFFs)\cite{herath2022} or PhaseD2NNs\cite{herath2022} as the optical feature extractors. We use SwinIR \cite{Liang2021SwinIRIR}, a state-of-the-art super-resolution pipeline, as the electronic reconstruction network. Our experiments with experimental datasets suggest that the proposed method can perform orders of magnitude faster QPM on cells than the current state of the art through $\times 64 - \times 256$ compression.

\begin{figure}[t]
    \centering
    \includegraphics[width=\textwidth]{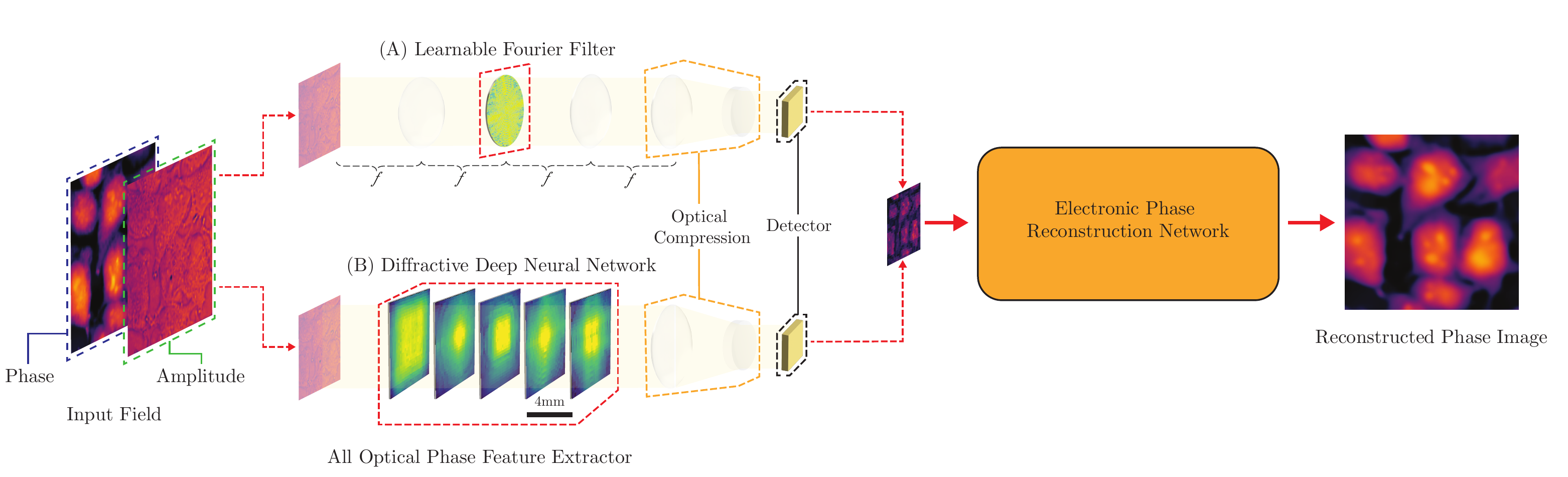}
    \caption{\textbf{End-to-end pipeline}: Input field with high contrast phase information is fed into the proposed optical phase feature extraction network. The resultant compressed output intensity field which contains the phase features is captured by the detector. The electronic phase reconstruction network utilizes these features to reconstruct phase information.}
    \label{fig:end_to_end}
\end{figure}

\section{Results}

\subsection{End-to-end Differentiable Compressive Quantitative Phase Microscopy}
\label{sec:results_intro}

We model the QPM system as a combination of an optical and electronic network that learns the task of reconstructing phase information of the input light field at the output of the electronic network. The optical network converts useful morphological features to an intensity field which is then captured by the detector array placed at the output plane of this network. Then, the electronic network constructs the phase map of the input light field from the captured features. This task is embedded in a loss function, and the entire network is parameterized in a differentiable manner. The parameters of the network are optimized to reduce the loss for a particular dataset. We follow a 3-stage optimization criteria for the improved stability of the end-to-end optimization; 1) optimize the all-optical phase extraction network; 2) optimize the electronic phase reconstruction network; 3) end-to-end fine-tuning. 

\paragraph{Optimize the all-optical phase extraction network.} Here the optical network is optimized to reconstruct the phase at its output intensity. For an input optical field $x_{in}= A_{in}e^{j\phi_{in}}$ we train an optical model $H_{O}$ through which the input field is propagated to produce the output field $x_{out} = A_{out}e^{j\phi_{out}} = H_{O}(x_{in})$. The  \textbf{phase reconstruction loss},  $\mathcal{L}_{\phi}$ introduced in previous work\cite{herath2022} is utilized here as,
\begin{equation}
    \label{eq:phase_recon1}
    \mathcal{L}_{\phi}=\mathbb{E}_{x_{in} \sim P_X}\left[L1(|A_{out}|^2, \phi_{in}/ (2\pi))\right],
\end{equation}
where, $P_X$ and $L1(.)$ respectively represent the probability distribution of phase objects and the L1 loss.

\paragraph{Optimize the electronic phase reconstruction network.} At this stage, we consider the end-to-end network, however, only the weights of the electronic network are optimized. The pretrained optical network discussed earlier is utilized as a feature extractor to encode the input phase. We demagnify the output field of the optical network to compress the intensity representation. The electronic super-resolution network reconstructs the input phase from the compressed intensity representation. The reconstructed phase information is given by $\hat{\phi} = H_{E}(D(|A_{out}|^2))$. Here $H_{E}(.)$ and $D(.)$ represent the electronic phase reconstruction network and the optical demagnification layer, respectively. $D(.)$, the optical demagnification layer is implemented through a stack of $2 \times 2$ average pooling operations \cite{Gholamalinezhad2020PoolingMI}. Similar to previous work \cite{Liang2021SwinIRIR}, we consider $\mathcal{L}_{swin}$, a combination of loss functions for this optimization,
\begin{equation}
    \label{eq:phase_recon2}
    \mathcal{L}_{swin}=\mathbb{E}_{x_{in} \sim P_X}\left[L1(\hat{\phi}, \phi_{in}/ (2\pi)) + \mathcal{L}_{perceptual}(\hat{\phi}, \phi_{in}/ (2\pi)) + \mathcal{L}_{adversarial}(\hat{\phi}, \phi_{in}/ (2\pi))\right],
\end{equation}
where, $\mathcal{L}_{perceptual}$ and $\mathcal{L}_{adversarial}$ represent the perceptual loss \cite{Liang2021SwinIRIR} and adversarial loss \cite{Liang2021SwinIRIR}, respectively.

\paragraph{End-to-end fine-tuning.} As the final stage, we finetune the entire optical-electronic network to reconstruct the phase at the output of the network. To improve the reconstruction in terms of capturing fine cell structures, we incorporate the negative structural similarity index measure (SSIM)\cite{wang2004} as the loss function.
\begin{equation}
    \label{eq:phase_recon3}
    \mathcal{L}_{SSIM}=\mathbb{E}_{x_{in} \sim P_X}- \frac{1}{M}\sum _{j=1}^{M}\frac{( 2\mu _{X_{j}} \mu _{Y_{j}} +C_{1})( 2\sigma _{X_{j}}{}_{Y_{j}} +C_{2})}{\left( \mu _{X_{j}}^{2} +\mu _{Y_{j}}^{2} +C_{1}\right)\left( \sigma _{X_{j}}^{2} +\sigma _{Y_{j}}^{2} +C_{2}\right)},
\end{equation}
$X_{j}$ and $Y_{j}$ represent equal-sized windows from a normalized input phase image ($\phi_{in}/{2\pi}$) and the corresponding reconstructed phase output ($\hat{\phi}$), respectively, for $M$ number of windows for an image. $P_X$ represents the probability distribution of input phase objects. $\mu_{X_{j}}, \mu_{Y_{j}}, \sigma_{X_{j}}, \sigma_{Y_{j}}, \sigma_{X_{j}Y_{j}}$ are the means, variances and the covariance of the $X_{j}$ and $Y_{j}$ windows, respectively. $C_{1} = (k_{1} \times L)^{2}$ and $C_{2} = (k_{2} \times L)^2$ are regularization parameters with $L = 1.0$, $k_{1} = 0.01$ and $k_{2} = 0.03$. 

In our formulation, since the input to the model is a complex-valued optical field, we only require input optical fields as training data. For the experiments, we utilize optical fields that are experimentally measured from phase objects. We consider a HeLa Cell dataset as our primary dataset. The optical fields in this dataset (i.e. full fields of view (full FoVs)) are $250 \mu m \times 250 \mu m$. Each full FoV is $789 \times 789$ pixel grid where each pixel is $316.4$ nm $\times 316.4$ nm. The dataset with full FoVs is divided into train and test sets. We cropped full FoVs into $256 \times 256$ sized patches (i.e. patch FoVs) for the end-to-end training of the proposed methods. Further details of the dataset are presented in the methods (section \ref{sec:datasets}). Using the above loss functions and training data, we train optical-electronic networks that consist of either an LFF or a PhaseD2NN (see section \ref{sec:results_end2end_lff}) as the optical network. To further improve the realisticity of the proposed method, we also conduct experiments with detector noise. In the next sections, we discuss these models and their results.

\begin{figure}[t!]
    \centering
    \includegraphics[width=0.9\textwidth]{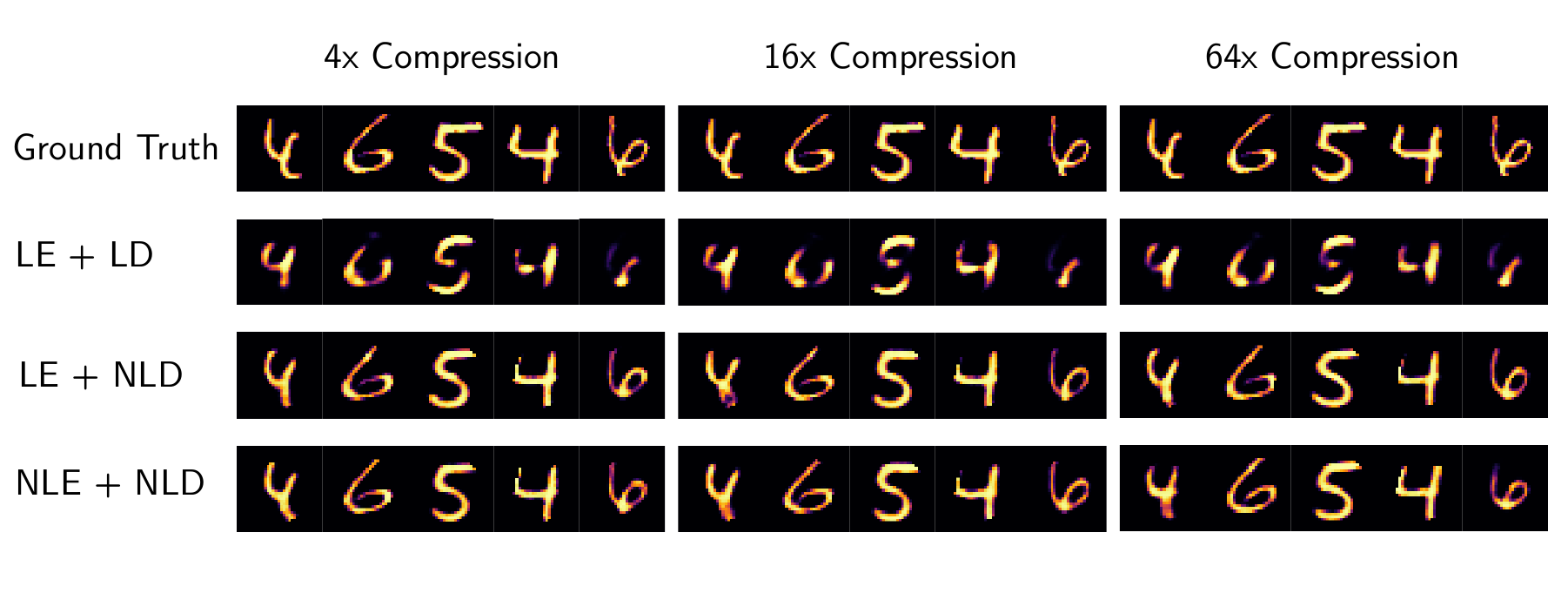}
    \caption{Compressibility of MNIST images using autoencoders (AE) with linear (L) and nonlinear (NL) encoder (E)/ decoder (D). LE, LD, NLE, and NLD represent linear encoder, linear decoder, non-linear encoder, and non-linear decoder respectively.}
    \label{fig:pre1}
\end{figure}

\subsection{Linear Encoding Does not Degrade Compressibility}
\label{sec:results_pre}

The optical feature extractor is a linear system. We therefore first established the feasibility of linear compression in comparison to nonlinear compressor models. 

\paragraph{Linear Encoding and Non-linear Decoding Allow Compression.} Prior to experimentation with an optical-electronic neural network, we first investigated the computational capabilities of the optics-based encoder and electronics-based decoder. Due to the linear nature of the optical encoder, we experimented on an autoencoder (AE) network \cite{autoencoder_review_paper} with a \textit{linear encoder} followed by a \textit{non-linear decoder}. The reconstruction results obtained from this network were compared with a fully linear autoencoder and a fully nonlinear autoencoder. The qualitative results in Fig. \ref{fig:pre1} show that the autoencoder network with a \textit{linear encoder} and \textit{non-linear decoder} performs on par with the fully nonlinear autoencoder.  

\begin{figure}[H]
    \centering
    \includegraphics[width=0.9\textwidth]{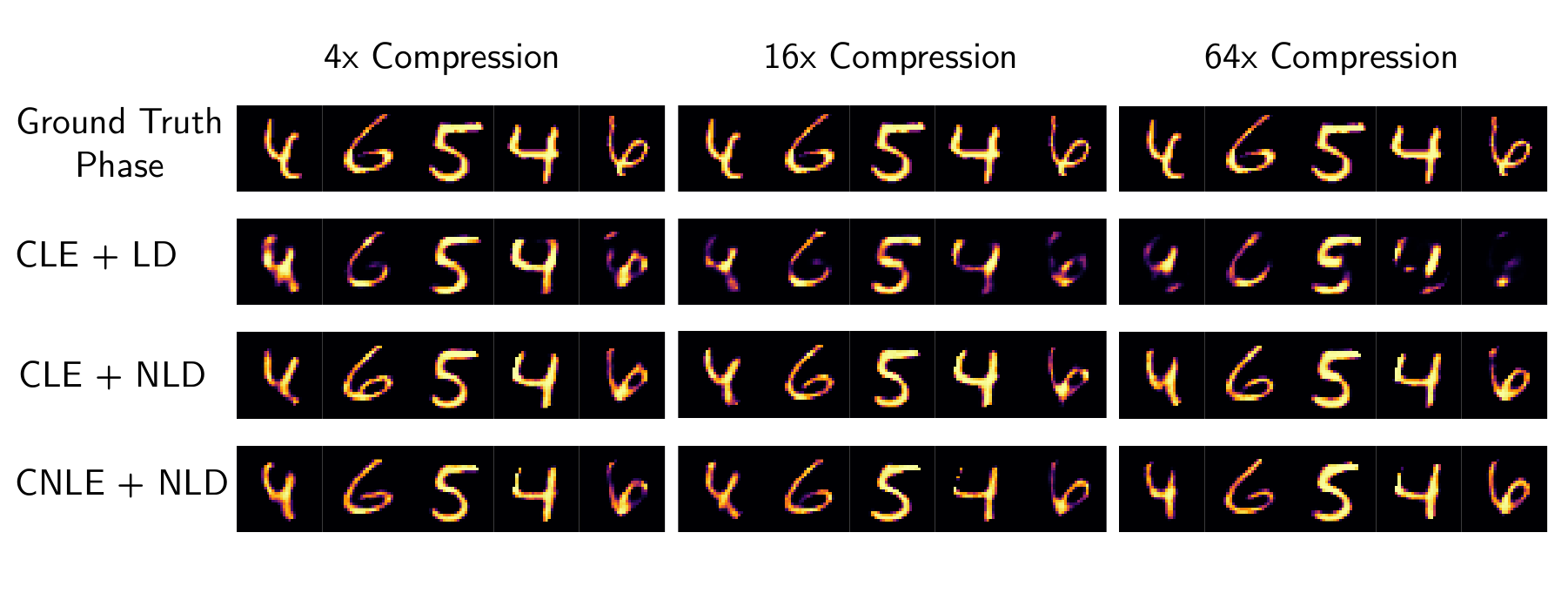}
    \caption{Phase to intensity conversion and compressibility of MNIST images using linear (L) and nonlinear (NL) encoder (E)/ decoder (D). Both the encoders are complex-valued hence denoted as CLE and CNLE.}
    \label{fig:pre2}
\end{figure}

\paragraph{Complex-valued Linear Encoding and Non-linear Decoding Allow Compression of Phase Information.} While we assessed the computational feasibility of a linear encoder followed by a non-linear decoder to perform reconstruction, in QPM, another main hurdle is that information of interest is in the phase of the light field. Therefore, we further assessed the ability of an autoencoder network (complex-valued linear encoder + non-linear decoder) to extract, compress, and reconstruct information encoded in the phase. Similar to previous results, Fig. \ref{fig:pre2} shows that a complex-valued linear encoder with a nonlinear decoder achieves similar qualitative performance as the complex-valued nonlinear encoder and nonlinear decoder.

These results suggest the computational feasibility of a linear optical network (encoder) followed by a nonlinear electronic network (decoder) to reconstruct information in the phase of the light field.

\begin{figure}[t]
    \centering
    \includegraphics[width=\textwidth]{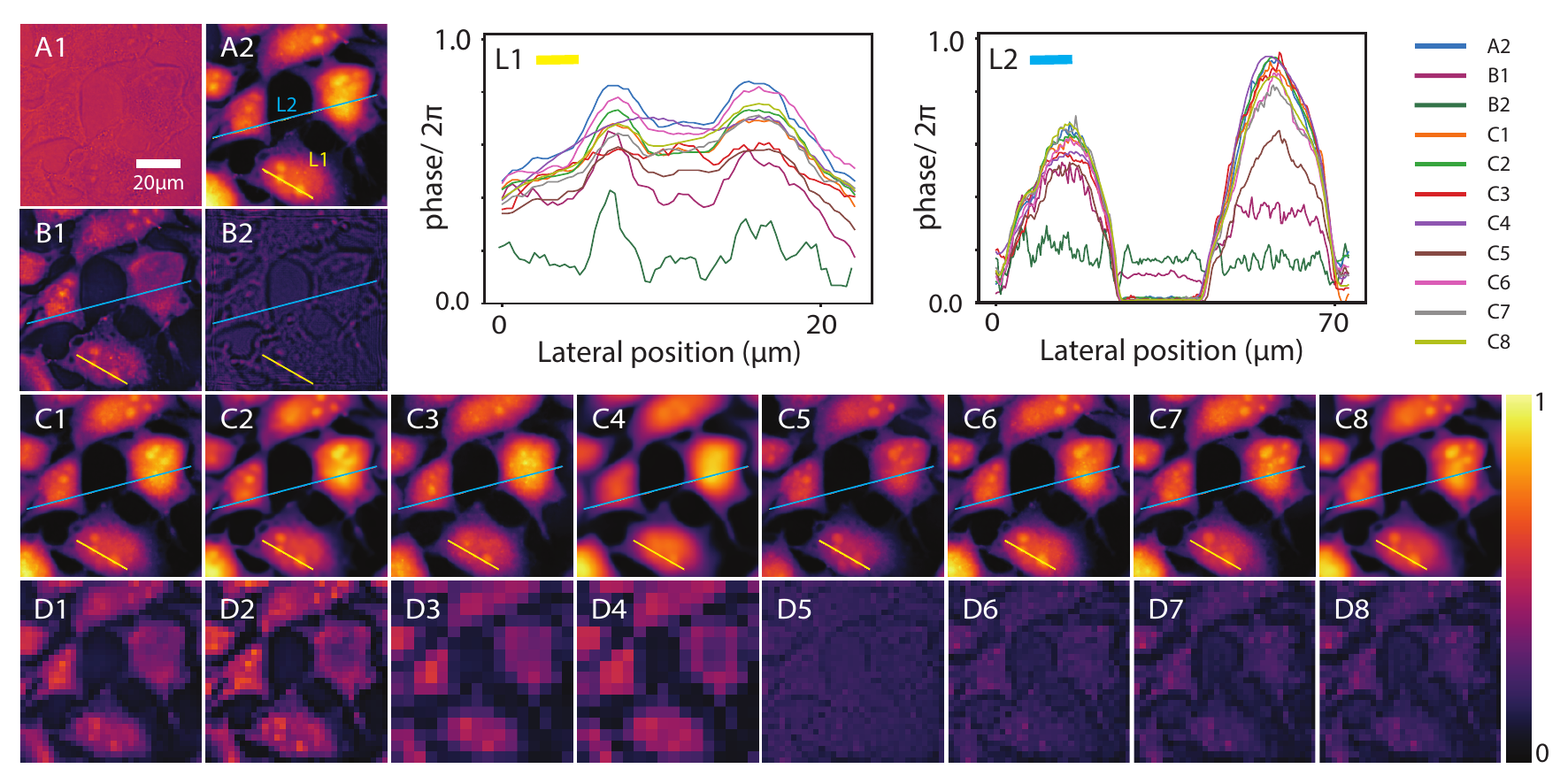}
    \caption{\textbf{Qualitative performance comparison}: Amplitude of input field (A1), phase of the input field (A2) (i.e. patch FoV) from the test set, all-optical phase to intensity conversion results \cite{herath2022} (i.e. baselines) using LFF (B1), PhaseD2NN (B2), Phase reconstructions from our approach 1: LFF + SwinIR with $\times 64$ compression without fine-tuning (C1), with fine-tuning (C2),
    LFF with $\times 256$ compression without fine-tuning (C3), with fine-tuning (C4), Phase reconstructions from our approach 2: PhaseD2NN + SwinIR with $\times 64$ compression without fine-tuning + 1 optical layer (C5), without fine-tuning + 3 optical layers (C6), without fine-tuning + 5 optical layers (C7), with fine-tuning + 5 optical layers (C8), Corresponding compressed output intensity fields of optical feature extractor (D1-8). Phase values along the L1 and L2 lines show the local and global resolving power of the proposed methods.}
    \label{fig:grid_results}
\end{figure}

\subsection{Optical Encoding and Electronic Decoding Enable Compressed QPM}

\label{sec:results_end2end_lff}
\label{sec:results_end2end_phased2nn}

Our results in section \ref{sec:results_pre} show that an autoencoder with a linear encoder and a non-linear decoder (\textit{AE:LE+NLD}) can reconstruct images as good as a fully nonlinear model. In this section, we propose our model with an all-optical feature extractor (linear encoder) and an electronic image reconstruction network (non-linear decoder) for QPM.

\paragraph{Learnable Fourier Filter (LFF) + SwinIR.} Based on previous work\cite{herath2022}, we first used a Learnable Fourier Filter (an LFF) as the optical feature extraction network. The LFF contained an optical $4$-$f$ system with a learnable circular Fourier filter. Similar to previous work \cite{herath2022}, the transmission coefficients of the circular Fourier filter were treated to be learnable. The input and output fields were $256 \times 256$  squared aperture grids. The circular Fourier filter had a radius of $128$ grid points. The coefficients of the filter were randomly initialized. We used SwinIR \cite{Liang2021SwinIRIR}, a state-of-the-art super-resolution network, as the electronic reconstruction network. We observed that directly training the end-to-end model (optical and electronic) was not ideal as the gradient flow between the optical and electronic networks was weak. Therefore, we employed the 3-stage criteria for the optimization of the end-to-end model (as discussed in section \ref{sec:results_intro}). We tested compression levels $\times 64$ and $\times 256$ for the compressed optical output intensity field in our experiments.

\begin{figure}[t!]
    \centering
    \includegraphics[width=\textwidth]{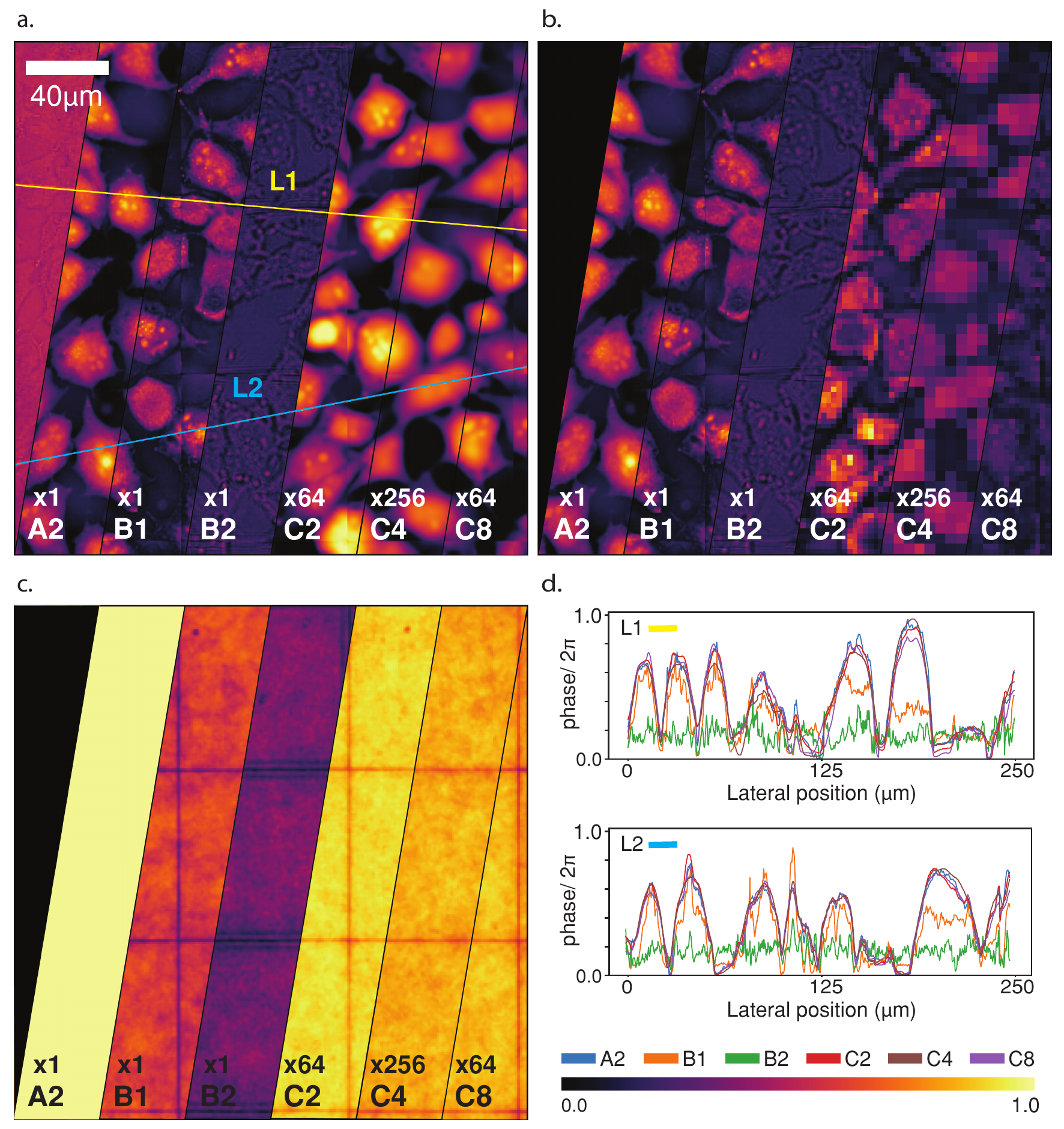}
    \caption{\textbf{Performance comparison of best methods}: a) The phase reconstructions. b) Compressed intensity fields at the detector. c) SSIM maps of reconstructions. d) The resolving power of the phase reconstructions is shown.  Phase of the input field (A2) of a full FoV from the test set, all-optical phase to intensity conversion results \cite{herath2022} (i.e. baselines) using LFF (B1), PhaseD2NN (B2), Phase reconstructions from our approach 1: LFF + SwinIR with $\times 64$ compression, with fine-tuning (C2), LFF with $\times 256$ compression with fine-tuning (C4), phase reconstructions from our approach 2: PhaseD2NN + SwinIR with $\times 64$ compression, with fine-tuning + 5 optical layers (C8).}
    
    \label{fig:main_results}
\end{figure}

Table \ref{tab:results} shows the performances at $\times 64$, $\times 256$ compression levels for the tested HeLa dataset (section \ref{sec:datasets}). For each compression level, performances are reported with and without the fine-tuning step. The corresponding qualitative results are shown in Figs. \ref{fig:grid_results} and \ref{fig:main_results}. All proposed methods outperformed all-optical baselines (B1, B2) with a significant margin in terms of SSIM (structural similarity index) and PSNR (peak signal-to-noise ratio) \cite{psnrvsssim}. End-to-end fine-tuning showed a considerable improvement in the performance for all the cases. Our best method achieved PSNR$= 29.76$ dB, SSIM$= 0.90$ performance at $\times 64$ compression, indicating that the proposed method is suitable for high-throughput QPM. Even at $\times 256$ compression, the proposed method outperformed all-optical baselines by a considerable margin with PSNR$= 27.62$ dB and SSIM$= 0.83$. We further tested our approach by including a noise model with Poisson noise and read noise \cite{2022udith}. We fine-tuned the best model (C2) with noise. A read noise with a standard deviation of $6.0$ and a detector maximum photon count of $10000$ were used. The proposed method with detector noise (E1) performed on par with the best model indicating that our LFF + SwinIR QPM is robust to real-world noise conditions. We further discuss the effect of the detector noise in the discussion (see section \ref{sec:discussion}).

\begin{table}
\small
\centering
\begin{tabular}{cclccccc}
\hline 
\multirow{2}{*}{\rotatebox{90}{Experiment}} & \multirow{2}{*}{\rotatebox{90}{Detector Noise}} & \multirow{2}{*}{\rotatebox{90}{Optical Net}} & \multirow{2}{*}{\rotatebox{90}{Compression}} & \multirow{2}{*}{\rotatebox{90}{Finetune}} & \multirow{2}{*}{\rotatebox{90}{\# layers}} & \multicolumn{2}{c}{\begin{tabular}[c]{@{}c@{}}\\ \\ \\ full FoV/ patch FoV \\ Metrics\end{tabular}} \\ [45pt] \cline{7-8} 
 &  &  &  &  &  & \textbf{PSNR} & \textbf{SSIM} \\ \hline
B1 & \multicolumn{5}{l}{All optical LFF} & \begin{tabular}[c]{@{}c@{}}16.1565/\\ 16.9761\end{tabular} & \begin{tabular}[c]{@{}c@{}}0.5880/\\ 0.6008\end{tabular} \\ \cline{2-8} 
B2 & \multicolumn{5}{l}{All optical PhaseD2NN} & \begin{tabular}[c]{@{}c@{}}12.3730/\\ 12.6631\end{tabular} & \begin{tabular}[c]{@{}c@{}}0.3163/\\ 0.3320\end{tabular} \\ \hline
C1 & \multirow{8}{*}{\xmark} & \multirow{4}{*}{LFF} & \multirow{2}{*}{64} & \xmark & - & \begin{tabular}[c]{@{}c@{}}23.8267/\\ 25.7840\end{tabular} & \begin{tabular}[c]{@{}c@{}}0.8225/\\ 0.8278\end{tabular} \\ \cline{5-8} 
C2 &  &  &  & \cmark & - & \textbf{\textcolor{red}{\begin{tabular}[c]{@{}c@{}}27.2579/\\ 29.7608\end{tabular}}} & \textbf{\textcolor{red}{\begin{tabular}[c]{@{}c@{}}0.8967/\\ 0.9031\end{tabular}}} \\ \cline{4-8} 
C3 &  &  & \multirow{2}{*}{256} & \xmark & - & \begin{tabular}[c]{@{}c@{}}22.5457/\\ 23.9536\end{tabular} & \begin{tabular}[c]{@{}c@{}}0.7470/\\ 0.7548\end{tabular} \\ \cline{5-8} 
C4 &  &  &  & \cmark & - & \begin{tabular}[c]{@{}c@{}}26.0003/\\ 27.6129\end{tabular} & \begin{tabular}[c]{@{}c@{}}0.8223/\\ 0.8302\end{tabular} \\ \cline{3-8} 
C5 &  & \multirow{4}{*}{\begin{tabular}[c]{@{}l@{}}Phase-\\ D2NN\end{tabular}} & \multirow{4}{*}{64} & \multirow{3}{*}{\xmark} & 1 & \begin{tabular}[c]{@{}c@{}}22.6495/\\ 23.8566\end{tabular} & \begin{tabular}[c]{@{}c@{}}0.7808/\\ 0.7889\end{tabular} \\ \cline{6-8} 
C6 &  &  &  &  & 3 & \begin{tabular}[c]{@{}c@{}}24.7560/\\ 26.0716\end{tabular} & \begin{tabular}[c]{@{}c@{}}0.8224/\\ 0.8313\end{tabular} \\ \cline{6-8} 
C7 &  &  &  &  & 5 & \begin{tabular}[c]{@{}c@{}}24.8015/\\ 26.0551\end{tabular} & \begin{tabular}[c]{@{}c@{}}0.8107/\\ 0.8185\end{tabular} \\ \cline{5-8} 
C8 &  &  &  & \cmark & 5 & \textbf{\textcolor{blue}{\begin{tabular}[c]{@{}c@{}}25.8617/\\ 27.2449\end{tabular}}} & \textbf{\textcolor{blue}{\begin{tabular}[c]{@{}c@{}}0.8519/\\ 0.8602\end{tabular}}} \\ \hline
E1 & \multirow{2}{*}{\cmark} & LFF & 64 & \cmark & - & \begin{tabular}[c]{@{}c@{}}27.3794/ \\ 29.8110\end{tabular} & \begin{tabular}[c]{@{}c@{}}0.8935/ \\ 0.8998\end{tabular} \\ \cline{3-8} 
E2 &  & \begin{tabular}[c]{@{}l@{}}Phase-\\ D2NN\end{tabular} & 64 & \cmark & 5 & \begin{tabular}[c]{@{}c@{}}25.7665/\\ 27.0651\end{tabular} & \begin{tabular}[c]{@{}c@{}}0.8477/\\ 0.8558\end{tabular} \\ \hline
\end{tabular}
\caption{\textbf{Performance comparison}: Best results for optical feature extraction networks \textbf{\textcolor{red}{LFF}, \textcolor{blue}{PhaseD2NN}} are highlighted. These best models are further fine-tuned end-to-end with the detector noise simulation (noise specifications of the detector: read noise standard deviation$=6.0$, maximum photon count$= 10000$) to improve realisticity. We calculate the patch and full FoV metrics on the test patch FoVs and full FoVs respectively. We reconstruct the full FoVs by tiling the reconstructed patch FoVs.}

\label{tab:results}
\end{table}

\paragraph{PhaseD2NN + SwinIR.} Second, we tested a PhaseD2NN \cite{herath2022} as the optical network in the proposed end-to-end framework. Similar to the previous section, the SwinIR super-resolution network was used for reconstruction. We selected the operating range of the PhaseD2NN as the visible wavelength ($\lambda = 632.8$ nm). on the same HeLa cell dataset (see section \ref{sec:datasets}).

The optical network consisted of $5$ optical layers each having $256 \times 256$ optical neural grid. The size of each neuron was $\frac{\lambda}{2} \times \frac{\lambda}{2}$ ($316.4$ nm $\times 316.4$ nm). Therefore, the size of the optical layer was $80.9984 \mu$m $\times 80.9984 \mu$m. Optical layers were separated with $3.373 \mu$m distance between each other. The distance between the input plane and the first optical layer was $3.373 \mu$m while the distance between the last optical layer and the detector plane was $5.904 \mu$m. Since the pixel size matched the PhaseD2NN neuron size, we could train the end-to-end network directly on the patch FoVs from the dataset. We followed the optimization criteria presented in section \ref{sec:results_intro} for the end-to-end training. Notably, we observed that in step 1, PhaseD2NN training was not stable due to the large number of physical parameters with a larger grid size (e.g., $256 \times 256$). To increase the stability and gradient flow of this optimization step, we used a sub-optimization-schedule (shown in Supplementary algorithm S1). We compressed the output intensity from the optical model $\times 64$ to obtain a higher throughput.

Table \ref{tab:results} shows the performances for $\times 64$ compression level. We report the performances while selecting different layers of the PhaseD2NN as the output layer. The corresponding learned power value was given by the algorithm S1 (Supplementary). The final model with $5$ layers was fine-tuned according to the proposed optimization steps. Similar to section \ref{sec:results_end2end_lff}, fine-tuning improved the performance. We explored different numbers of diffractive layers for the PhaseD2NN without the fine-tuning step and the results are presented in Table \ref{tab:results}.

We performed further experiments with the $5$ layer PhaseD2NN (C8 and E2). Our method achieved the best performance of PSNR$= 27.24$ dB, SSIM$= 0.86$ with $\times 64$ compression which was considerably higher than the all-optical baselines. Similar to the previous section, we injected detector noise with similar specifications of a maximum photon count of $10000$ and detector read noise standard deviation of $6.0$. The resultant performance with the detector noise (E2) was on par with the best model without the noise (C8). This indicates that our PhaseD2NN + SwinIR QPM is robust to real-world noise conditions.

\section{Discussion}
\label{sec:discussion}

\paragraph{Overall Comparison.} Fig. \ref{fig:main_results} presents the qualitative results for best-performing models. Fig. \ref{fig:main_results}(d) shows that the proposed differentiable optical-electronic QPM systems have a higher resolving capability compared to all-optical baselines.  Fig. \ref{fig:main_results}(c) SSIM maps show how our methods perform for different regions of full field-of-view (FoV). Low SSIM in edges indicates that there is room to improve the proposed QPM just by refining the edges of generated patches. We observed that, even though both proposed methods: i.e, LFF with SwinIR; and PhaseD2NN with SwinIR outperformed the all-optical baselines by a considerable margin, the LFF-based method performed better than the PhaseD2NN-based one. Further studies are needed to investigate the reason for this behavior. 

\paragraph{Stability of PhaseD2NN Training.} We observed that the optimization step 1, i.e., all-optical reconstruction (see section \ref{sec:results_intro}), is not stable for the PhaseD2NN. We suspect that the reason for this instability is the large FoV (of $256 \times 256$) resulting in a large number of learnable parameters. To overcome this, we used a sub-optimization-schedule for the PhaseD2NN training motivated by progressive growing learning principles \cite{2018pgGAN} (see algorithm S1 in the supplementary). Instead of training the PhaseD2NN in an end-to-end fashion, here we optimize the PhaseD2NN layer by layer progressively with the phase reconstruction loss. With this schedule, we could efficiently train the optical network. Even though one can argue that the proposed schedule leads to a sub-optimal solution, we achieved a sufficient performance for QPM \cite{Wang2019_ynet} with this schedule. Nevertheless, an interesting future direction is to explore more efficient methods to train large D2NNs.

\paragraph{Resolving Power of Differentiable Optical-Electronic QPM.} We analyzed the resolving power of the proposed differentiable optical-electronic QPM in Fig. \ref{fig:grid_results}(d). The plots demonstrate the phase value variations along lines L1 and L2 on the patch FoV. Phase variations along L1 and L2 further show the superior resolving ability of the proposed methods for local and global features respectively. Furthermore, fine-tuning improved the resolving capability for most of the models. 

\paragraph{Effect of Photodetector Noise.} To further evaluate the behavior of the proposed method with detector noise, we evaluated the method with maximum photon counts of $100$ and $10000$, and read noise standard deviations of $4.0$ and $6.0$. Table \ref{tab:noise_exps} shows that differentiable optical-electronic QPM is robust to noise when the maximum photon count is $10000$ (for most QPM applications high light conditions can be used). Even though the proposed method performs worse for lower photon counts (e.g., $100$), it still performs far better than all-optical baselines. We further note that the performance degradation due to higher read noise is insignificant for larger photon counts (e.g., $10000$).

\paragraph{Compressibility limitations.} Last, we tested our LFF-based approach on a QPM dataset of tissue with much more complex features (see section \ref{sec:datasets}). The goal of this experiment was to investigate the limitations of our approach at high compression levels. We observed that our method failed to reconstruct high-resolution features at both $\times64$ and $\times256$ (see supplementary Fig. S1). There could be two potential reasons for the subpar performance. It could be the case that the optical compressor cannot efficiently convert phase information to the latent intensity field at the detector. Alternatively, it could be the case that the reconstruction network is not capable of reconstructing highly compressed information from images with complex features. To investigate the latter we tested our reconstruction network on a simple resolution enhancement task on the same tissue dataset. As shown in supplementary Fig. S1, here too the reconstruction network failed. Thus we conclude that in our method, the compressibility is limited in the presence of complex features. Further studies are required to establish the compressibility bounds for data distributions of interest.

\begin{table}
\centering
\begin{tabular}{lcccc}
\hline
\multirow{2}{*}{Optical Net} & \multicolumn{2}{c}{\begin{tabular}[c]{@{}c@{}}Noise \\ Specifications\end{tabular}} & \multicolumn{2}{c}{\begin{tabular}[c]{@{}c@{}}full FoV/ patch FoV \\ metrics\end{tabular}} \\ \cline{2-5} 
 & \begin{tabular}[c]{@{}c@{}}max. \\ photon\\ count\end{tabular} & $\sigma_{read}$ & \textbf{PSNR} & \textbf{SSIM} \\ \hline
\multirow{4}{*}{LFF} & 100 & 4 & \begin{tabular}[c]{@{}c@{}}23.4928/\\ 25.0266\end{tabular} & \begin{tabular}[c]{@{}c@{}}0.777/\\ 0.7834\end{tabular} \\ \cline{3-5} 
 &  & 6 & \begin{tabular}[c]{@{}c@{}}22.5004/\\ 24.1487\end{tabular} & \begin{tabular}[c]{@{}c@{}}0.7606/\\ 0.7663\end{tabular} \\ \cline{2-5} 
 & 10000 & 4 & \textbf{\begin{tabular}[c]{@{}c@{}}27.4122/\\ 29.8110\end{tabular}} & \textbf{\begin{tabular}[c]{@{}c@{}}0.8935/\\ 0.8997\end{tabular}} \\ \cline{3-5} 
 &  & 6 & \textbf{\begin{tabular}[c]{@{}c@{}}27.3794/\\ 29.8110\end{tabular}} & \textbf{\begin{tabular}[c]{@{}c@{}}0.8935/\\ 0.8998\end{tabular}} \\ \hline
\multirow{4}{*}{\begin{tabular}[c]{@{}l@{}}Phase-\\ D2NN\end{tabular}} & 100 & 4 & \begin{tabular}[c]{@{}c@{}}17.8942/\\ 18.7526\end{tabular} & \begin{tabular}[c]{@{}c@{}}0.6441/\\ 0.6502\end{tabular} \\ \cline{3-5} 
 &  & 6 & \begin{tabular}[c]{@{}c@{}}16.8953/\\ 17.7111\end{tabular} & \begin{tabular}[c]{@{}c@{}}0.6094/\\ 0.6158\end{tabular} \\ \cline{2-5} 
 & 10000 & 4 & \textbf{\begin{tabular}[c]{@{}c@{}}25.7193/\\ 27.0456\end{tabular}} & \textbf{\begin{tabular}[c]{@{}c@{}}0.8478/\\ 0.8559\end{tabular}} \\ \cline{3-5} 
 &  & 6 & \textbf{\begin{tabular}[c]{@{}c@{}}25.7665/\\ 27.0651\end{tabular}} & \textbf{\begin{tabular}[c]{@{}c@{}}0.8477/\\ 0.8558\end{tabular}} \\ \hline
\end{tabular}
\caption{\textbf{Performance of our method for different detector noise conditions}: Our best models (C2, C8 in table \ref{tab:results}) are further fine-tuned with the corresponding noise specifications.}
\label{tab:noise_exps}
\end{table}

\section{Conclusion}

Quantitative phase microscopy (QPM) is an emerging label-free imaging modality with a wide range of biological and clinical applications. Recent advances in QPM are focused on developing fast instruments through better detectors and fast deep-learning-based inverse solvers. However, currently, the QPM throughput is fundamentally limited by the pixel throughput of the imaging detectors. Orthogonal to current advances, to improve QPM throughput beyond the hardware bottleneck, here we propose to use content-aware compressive data acquisition. Specifically, we utilize learnable optical front-ends to extract compressed phase features. A state-of-the-art transformer deep network then decodes the captured information to quantitatively reconstruct the phase image. The proposed pipeline inherently improves the imaging speed while achieving high-quality reconstructions. Moreover, the advances presented in this work can lead to similar developments in a wide range of label-free coherent imaging modalities such as photothermal, coherent anti-Stokes Raman scattering (CARS), and stimulated Raman scattering (SRS).

\section{Methods}

\label{sec:datasets}

\subsection{Datasets}

In our numerical experiments, we used two datasets.

\paragraph{HeLa Cell Dataset :}We used a HeLa cell dataset \cite{herath2022} as the primary dataset for our experiments. We followed the sample preparation procedure explained in previous work \cite{herath2022}. The initial dataset contained 501 complex-valued images (i.e. detected FoVs). Each detected FoV was obtained by a camera with a $2304 \times 2304$ pixel grid where the pixel size was $6.5 \ \mu\text{m} \times 6.5 \ \mu\text{m}$. The light field from the specimen was magnified $\times 60$ before imaging onto the detector. 

To pre-process the dataset, we first calculated the side length of the light fields before the magnification ($= \frac{2304 pixels \times 6.5 \ \mu\text{m} /pixel}{60} = 249.6 \ \mu\text{m}$). Second, we calculated the number of $316.4$ nm $\times 316.4$ nm sized pixels in these light fields ($= round(\frac{249.6 \ \mu\text{m}}{316.4 \text{ nm}/ pixel} = 789 pixels)$). Finally, we resized the detected FoVs (i.e. $2304 \times 2304$ pixel grids) into $789 \times 789$ pixel grids. This resulted in the light field before the magnification with a pixel size of $316.4$ nm $\times 316.4$ nm. We refer to these FoVs as full FoVs. We obtained train and test sets by dividing the full FoV dataset into $401$ and $100$ sets. For the training of the proposed networks, we used $256 \times 256$ cropped patches (i.e. patch FoVs) from the full FoVs.

\paragraph{Tissue Dataset:}We also acquired a tissue dataset to further validate our observations and to derive an empirical upper bound for the results. We followed preparation, acquisition, and preprocessing procedures similar to HeLa cells, with a magnification of $\times 20$. There were 470 detected FoVs. Camera had $2367 \times 2367$ pixel grid where the pixel size was $6.5 \ \mu\text{m} \times 6.5 \ \mu\text{m}$. Side length of the light fields before the magnification was, $\frac{2304 pixels \times 6.5 \ \mu\text{m} /pixel}{20} = 748.8 \ \mu\text{m}$). Number of $316.4$ nm $\times 316.4$ nm sized pixels in these light field was $= round(\frac{748.8 \ \mu\text{m}}{316.4 \text{ nm}/ pixel} = 2367 pixels)$. We resized the detected FoVs (i.e. $2304 \times 2304$ pixel grids) into $2367 \times 2367$ pixel grids to match the pixel sizes of the light fields and the algorithm (full FoVs). The full FoV dataset was divided into $470$ and $117$, train and test sets.

\subsection{Implementation Details}

We implemented the proposed optical-electronic networks with Python version 3.6.13. We used auto differentiation in PyTorch \cite{pytorch} framework version 1.8.0 for the joint optimization/ training of the proposed optical-electronic networks. All experiments were conducted on a server with 12 Intel(R) Xeon(R) Platinum 8358 (2.60 GHz) CPU Cores, an NVIDIA A100 Graphics Processing Unit with 40 GB memory running on the Centos operating system.

We used batch size 32, learning rates of 0.1, 0.001 respectively for LFF and PhaseD2NN in the optimization stage 1. LFF was trained for 1500 epochs with multi-step learning rate scheduler \cite{pytorch} (milestones : [50, 400, 650, 1000, 1400], $\gamma= 0.1$). PhaseD2NN was trained for 1500 epochs after each optimizer initialization step in algorithm S1. For joint multi-layer optimizations in algorithm S1, a learning rate of 0.00005 was used for better stability. For the optimization stage 2, we followed similar training configurations used in SwinIR \cite{Liang2021SwinIRIR}. Lastly, for the final optimization stage (i.e. end-to-end fine-tuning), we fine-tuned the LFF + SwinIR and PhaseD2NN + SwinIR for 24000 and 3000 epochs with a learning rate of $5\times 10^{-6}$, respectively. We used Adam \cite{adam} as the optimizer for all optimizations.

\section{Backmatter}

\begin{backmatter}
\bmsection{Funding}
The Research Council of Norway; National Institute of Biomedical Imaging and Bioengineering (1-R21-MH130067-01).

\bmsection{Acknowledgments}
This work was supported by the Center for Advanced Imaging at Harvard University (D.N.W., U.H., K.H., R.H., and H.K.), 1-R21-MH130067-01(U.H., D.N.W). D.N.W. is also supported by the John Harvard Distinguished Science Fellowship Program within the FAS Division of Science of Harvard University. 

\bmsection{Disclosures}
The authors declare that there are no conflicts of interest related to this article

\bmsection{Data availability} Data underlying the results presented in this paper can be obtained from the authors upon reasonable request.

\bmsection{Supplemental document}
See the supplemental document for supporting content.

\end{backmatter}

\bibliography{refs}

\begin{thebibliography}{10}
\newcommand{\enquote}[1]{``#1''}

\bibitem{3popescu2006}
G.~Popescu, T.~Ikeda, R.~R. Dasari, and M.~S. Feld, \enquote{Diffraction phase
  microscopy for quantifying cell structure and dynamics,}
  {\protect\JournalTitle{Opt. Lett.}} \textbf{31}, 775--777 (2006).

\bibitem{4park2006}
Y.~Park, G.~Popescu, K.~Badizadegan, R.~R. Dasari, and M.~S. Feld,
  \enquote{Diffraction phase and fluorescence microscopy,}
  {\protect\JournalTitle{Opt. Express}} \textbf{14}, 8263--8268 (2006).

\bibitem{5fang-yen2007}
C.~Fang-Yen, S.~Oh, Y.~Park, W.~Choi, S.~Song, H.~S. Seung, R.~R. Dasari, and
  M.~S. Feld, \enquote{Imaging voltage-dependent cell motions with heterodyne
  mach-zehnder phase microscopy,} {\protect\JournalTitle{Opt. Lett.}}
  \textbf{32}, 1572--1574 (2007).

\bibitem{6amin2007}
M.~S. Amin, Y.~Park, N.~Lue, R.~R. Dasari, K.~Badizadegan, M.~S. Feld, and
  G.~Popescu, \enquote{Microrheology of red blood cell membranes using dynamic
  scattering microscopy,} {\protect\JournalTitle{Opt. Express}} \textbf{15},
  17001--17009 (2007).

\bibitem{7sung2014}
Y.~Sung, N.~Lue, B.~Hamza, J.~Martel, D.~Irimia, R.~R. Dasari, W.~Choi,
  Z.~Yaqoob, and P.~So, \enquote{Three-dimensional holographic refractive-index
  measurement of continuously flowing cells in a microfluidic channel,}
  {\protect\JournalTitle{Phys. Rev. Applied}} \textbf{1}, 014002 (2014).

\bibitem{8choi2007}
W.~Choi, C.~Fang-Yen, K.~Badizadegan, S.~Oh, N.~Lue, R.~R. Dasari, and M.~S.
  Feld, \enquote{Tomographic phase microscopy,} {\protect\JournalTitle{Nat
  Methods}} \textbf{4}, 717--719 (2007).

\bibitem{Park2018}
Y.~K. Park, C.~Depeursinge, and G.~Popescu, \enquote{{Quantitative phase
  imaging in biomedicine},} {\protect\JournalTitle{Nature Photonics}}
  \textbf{12}, 578--589 (2018).

\bibitem{10jo2017}
Y.~Jo, S.~Park, J.~Jung, J.~Yoon, H.~Joo, M.-H. Kim, S.-J. Kang, M.~C. Choi,
  S.~Y. Lee, and Y.~Park, \enquote{Holographic deep learning for rapid optical
  screening of anthrax spores,} {\protect\JournalTitle{Sci Adv}} \textbf{3},
  e1700606 (2017).

\bibitem{Roitshtain2017}
D.~Roitshtain, L.~Wolbromsky, E.~Bal, H.~Greenspan, L.~L. Satterwhite, and
  N.~T. Shaked, \enquote{{Quantitative phase microscopy spatial signatures of
  cancer cells},} {\protect\JournalTitle{Cytometry Part A}} \textbf{91},
  482--493 (2017).

\bibitem{12majeed2019}
H.~Majeed, A.~Keikhosravi, M.~Kandel, T.~Nguyen, Y.~Liu, A.~Kajdacsy-Balla,
  K.~Tangella, K.~Eliceiri, and G.~Popescu, \enquote{Quantitative
  histopathology of stained tissues using color spatial light interference
  microscopy (cslim),} {\protect\JournalTitle{Scientific Reports}} \textbf{9}
  (2019).

\bibitem{Rivenson2019PhaseStain:Learning}
Y.~Rivenson, T.~Liu, Z.~Wei, Y.~Zhang, K.~de~Haan, and A.~Ozcan,
  \enquote{{PhaseStain}: The digital staining of label-free quantitative phase
  microscopy images using deep learning,} {\protect\JournalTitle{Light: Science
  and Applications}} \textbf{8} (2019).

\bibitem{14Wang2011}
Z.~Wang, K.~Tangella, A.~Balla, and G.~Popescu, \enquote{Tissue refractive
  index as marker of disease,} {\protect\JournalTitle{J Biomed Opt}}
  \textbf{16}, 116017 (2011).

\bibitem{15Cree2014}
I.~A. Cree, Z.~Deans, M.~J.~L. Ligtenberg, N.~Normanno, A.~Edsj{\"o},
  E.~Rouleau, F.~Sol{\'e}, E.~Thunnissen, W.~Timens, E.~Schuuring, E.~Dequeker,
  S.~Murray, M.~Dietel, P.~Groenen, J.~H. Van~Krieken, {European Society of
  Pathology Task Force on Quality Assurance in Molecular Pathology}, and {Royal
  College of Pathologists}, \enquote{Guidance for laboratories performing
  molecular pathology for cancer patients,} {\protect\JournalTitle{J Clin
  Pathol}} \textbf{67}, 923--931 (2014).

\bibitem{16Kandel2019}
M.~Kandel, C.~Hu, G.~Naseri~Kouzehgarani, E.~Min, K.~Sullivan, H.~Kong, J.~Li,
  D.~Robson, M.~Gillette, C.~Best-Popescu, and G.~Popescu,
  \enquote{Epi-illumination gradient light interference microscopy for imaging
  opaque structures,} {\protect\JournalTitle{Nature Communications}}
  \textbf{10} (2019).

\bibitem{17sung2012}
Y.~Sung, W.~Choi, N.~Lue, R.~R. Dasari, and Z.~Yaqoob, \enquote{Stain-free
  quantification of chromosomes in live cells using regularized tomographic
  phase microscopy,} {\protect\JournalTitle{PLoS One}} \textbf{7}, e49502
  (2012).

\bibitem{Sung2013SizeMicroscopy}
Y.~Sung, A.~Tzur, S.~Oh, W.~Choi, V.~Li, R.~R. Dasari, Z.~Yaqoob, and M.~W.
  Kirschner, \enquote{{Size homeostasis in adherent cells studied by synthetic
  phase microscopy},} {\protect\JournalTitle{Proceedings of the National
  Academy of Sciences of the United States of America}} \textbf{110} (2013).

\bibitem{19Hosseini2016}
P.~Hosseini, S.~Z. Abidi, E.~Du, D.~P. Papageorgiou, Y.~Choi, Y.~Park, J.~M.
  Higgins, G.~J. Kato, S.~Suresh, M.~Dao, Z.~Yaqoob, and P.~T.~C. So,
  \enquote{Cellular normoxic biophysical markers of hydroxyurea treatment in
  sickle cell disease,} {\protect\JournalTitle{Proc Natl Acad Sci U S A}}
  \textbf{113}, 9527--9532 (2016).

\bibitem{20Singh2019}
V.~Singh, Y.~A. Yang, H.~Yu, R.~Kamm, Z.~Yaqoob, and P.~So, \enquote{Studying
  nucleic envelope and plasma membrane mechanics of eukaryotic cells using
  confocal reflectance interferometric microscopy,}
  {\protect\JournalTitle{Nature Communications}} \textbf{10} (2019).

\bibitem{Zernike1942}
F.~Zernike, \enquote{Observation of transparent objects,}
  {\protect\JournalTitle{Physica}} pp. 974--986 (1942).

\bibitem{Gluckstad1996}
J.~Gl{\"{u}}ckstad, \enquote{{Phase contrast image synthesis},}
  {\protect\JournalTitle{Optics Communications}} \textbf{130}, 225--230 (1996).

\bibitem{diffphasecontrast}
N.~{Shibata}, S.~D. {Findlay}, Y.~{Kohno}, H.~{Sawada}, Y.~{Kondo}, and
  Y.~{Ikuhara}, \enquote{{Differential phase-contrast microscopy at atomic
  resolution},} {\protect\JournalTitle{Nature Physics}} \textbf{8}, 611--615
  (2012).

\bibitem{Jo2018}
Y.~J. Jo, H.~Cho, S.~Y. Lee, G.~Choi, G.~Kim, H.~S. Min, and Y.~K. Park,
  \enquote{{Quantitative phase imaging and artificial intelligence: A review},}
  {\protect\JournalTitle{IEEE Journal of Selected Topics in Quantum
  Electronics}} \textbf{25} (2018).

\bibitem{Kim2013}
K.~Kim, K.~S. Kim, H.~Park, J.~C. Ye, and Y.~Park, \enquote{Real-time
  visualization of 3-d dynamic microscopic objects using optical diffraction
  tomography,} {\protect\JournalTitle{Opt. Express}} \textbf{21}, 32269--32278
  (2013).

\bibitem{Lim2015}
J.~Lim, K.~Lee, K.~H. Jin, S.~Shin, S.~Lee, Y.~Park, and J.~C. Ye,
  \enquote{Comparative study of iterative reconstruction algorithms for missing
  cone problems in optical diffraction tomography,} {\protect\JournalTitle{Opt.
  Express}} \textbf{23}, 16933--16948 (2015).

\bibitem{Sung2009}
Y.~Sung, W.~Choi, C.~Fang-Yen, K.~Badizadegan, R.~R. Dasari, and M.~S. Feld,
  \enquote{{Optical diffraction tomography for high resolution live cell
  imaging},} {\protect\JournalTitle{Optics InfoBase Conference Papers}}
  \textbf{17}, 1977--1979 (2009).

\bibitem{Nguyen2018}
T.~Nguyen, V.~Bui, and G.~Nehmetallah, \enquote{Computational optical
  tomography using 3d deep convolutional neural networks (3d-dcnns),}
  {\protect\JournalTitle{Optical Engineering}} \textbf{57} (2018).

\bibitem{Di2020}
J.~Di, K.~Wang, Y.~Li, and J.~Zhao, \enquote{Deep learning-based holographic
  reconstruction in digital holography,} in \emph{Imaging and Applied Optics
  Congress,}  (Optica Publishing Group, 2020), p. HTu4B.2.

\bibitem{Zhu2021}
Y.~Zhu, C.~H. Yeung, and E.~Y. Lam, \enquote{Digital holographic imaging and
  classification of microplastics using deep transfer learning,}
  {\protect\JournalTitle{Appl. Opt.}} \textbf{60}, A38--A47 (2021).

\bibitem{Wang2020_y4net}
K.~Wang, Q.~Kemao, J.~Di, and J.~Zhao, \enquote{Y4-net: A deep learning
  solution to one-shot dual-wavelength digital holographic reconstruction,}
  {\protect\JournalTitle{Opt. Lett.}} \textbf{45}, 4220--4223 (2020).

\bibitem{Wang2018_eholonet}
H.~Wang, M.~Lyu, and G.~Situ, \enquote{eholonet: A learning-based end-to-end
  approach for in-line digital holographic reconstruction,}
  {\protect\JournalTitle{Opt. Express}} \textbf{26}, 22603--22614 (2018).

\bibitem{Wang2019_ynet}
K.~Wang, J.~Dou, Q.~Kemao, J.~Di, and J.~Zhao, \enquote{Y-net: A one-to-two
  deep learning framework for digital holographic reconstruction,}
  {\protect\JournalTitle{Opt. Lett.}} \textbf{44}, 4765--4768 (2019).

\bibitem{Kellman2019Physics-BasedImaging}
M.~R. Kellman, E.~Bostan, N.~A. Repina, and L.~Waller, \enquote{{Physics-based
  learned design: Optimized coded-illumination for quantitative phase
  imaging},} {\protect\JournalTitle{IEEE Transactions on Computational
  Imaging}} \textbf{5} (2019).

\bibitem{Matlock:19}
A.~Matlock and L.~Tian, \enquote{High-throughput, volumetric quantitative phase
  imaging with multiplexed intensity diffraction tomography,}
  {\protect\JournalTitle{Biomed. Opt. Express}} \textbf{10}, 6432--6448 (2019).

\bibitem{21yeh1995}
P.~Yeh and C.~Gu, \emph{Landmark Papers on Photorefractive Nonlinear Optics}
  (World Scientific, 1995).

\bibitem{23Vincent2012}
V.~Studer, J.~Bobin, M.~Chahid, H.~S. Mousavi, E.~Candes, and M.~Dahan,
  \enquote{Compressive fluorescence microscopy for biological and hyperspectral
  imaging,} {\protect\JournalTitle{Proceedings of the National Academy of
  Sciences}} \textbf{109}, E1679--E1687 (2012).

\bibitem{2022udith}
U.~Haputhanthri, A.~Seeber, and D.~Wadduwage, \enquote{Differentiable
  microscopy for content and task aware compressive fluorescence imaging,}
  (2022).

\bibitem{herath2022}
K.~Herath, U.~Haputhanthri, R.~Hettiarachchi, H.~Kariyawasam, A.~Ahmad, B.~S.
  Ahluwalia, C.~U.~S. Edussooriya, and D.~Wadduwage, \enquote{Differentiable
  microscopy designs an all optical quantitative phase microscope,}  (2022).

\bibitem{Liang2021SwinIRIR}
J.~Liang, J.~Cao, G.~Sun, K.~Zhang, L.~V. Gool, and R.~Timofte,
  \enquote{Swinir: Image restoration using swin transformer,}
  {\protect\JournalTitle{2021 IEEE/CVF International Conference on Computer
  Vision Workshops (ICCVW)}} pp. 1833--1844 (2021).

\bibitem{Gholamalinezhad2020PoolingMI}
H.~Gholamalinezhad and H.~Khosravi, \enquote{Pooling methods in deep neural
  networks, a review,} {\protect\JournalTitle{ArXiv}} \textbf{abs/2009.07485}
  (2020).

\bibitem{wang2004}
Z.~Wang, A.~Bovik, H.~Sheikh, and E.~Simoncelli, \enquote{Image quality
  assessment: From error visibility to structural similarity,}
  {\protect\JournalTitle{IEEE Transactions on Image Processing}} \textbf{13},
  600--612 (2004).

\bibitem{autoencoder_review_paper}
Y.~Wang, H.~Yao, and S.~Zhao, \enquote{Auto-encoder based dimensionality
  reduction,} {\protect\JournalTitle{Neurocomputing}} \textbf{184}, 232--242
  (2016). RoLoD: Robust Local Descriptors for Computer Vision 2014.

\bibitem{psnrvsssim}
A.~Horé and D.~Ziou, \enquote{Image quality metrics: {PSNR} vs. {SSIM},} in
  \emph{2010 20th International Conference on Pattern Recognition,}  (2010),
  pp. 2366--2369.

\bibitem{2018pgGAN}
T.~Karras, T.~Aila, S.~Laine, and J.~Lehtinen, \enquote{Progressive growing of
  gans for improved quality, stability, and variation,}
  {\protect\JournalTitle{ArXiv}} \textbf{abs/1710.10196} (2018).

\bibitem{pytorch}
A.~Paszke, S.~Gross, F.~Massa, A.~Lerer, J.~Bradbury, G.~Chanan, T.~Killeen,
  Z.~Lin, N.~Gimelshein, L.~Antiga, A.~Desmaison, A.~Kopf, E.~Yang, Z.~DeVito,
  M.~Raison, A.~Tejani, S.~Chilamkurthy, B.~Steiner, L.~Fang, J.~Bai, and
  S.~Chintala, \enquote{Pytorch: An imperative style, high-performance deep
  learning library,} in \emph{Advances in Neural Information Processing Systems
  32,}  H.~Wallach, H.~Larochelle, A.~Beygelzimer, F.~d\textquotesingle
  Alch\'{e}-Buc, E.~Fox, and R.~Garnett, eds. (Curran Associates, Inc., 2019),
  pp. 8024--8035.

\bibitem{adam}
D.~Kingma and J.~Ba, \enquote{Adam: A method for stochastic optimization,}
  {\protect\JournalTitle{International Conference on Learning Representations}}
   (2014).

\end{thebibliography}

\end{document}


\maketitle

\section{Sub-optimization-schedule for stable all-optical PhaseD2NN training}

We observe that using PhaseD2NN as the optical network hinders the convergence of the optimization step-1 (section 2.1 in the manuscript). To improve the stability of this step, we consider a sub-optimization-schedule in algorithm \ref{algo:phased2nn-sub}.

\RestyleAlgo{ruled}
\SetKwComment{Comment}{/* }{ */}
\begin{algorithm}[H]
\caption{Sub-optimization-schedule for stable all-optical PhaseD2NN training}\label{algo:phased2nn-sub}
\KwData{$P_X$}
\KwResult{$H_O^*(.)$ \Comment*[r]{Optimal PhaseD2NN all-optical model}
}

$H_O(.) \gets (f_n \circ f_{n-1} \circ ... \circ f_1) (.)$ \Comment*[r]{Define PhaseD2NN. $f_n$ is the $n^{th}$ layer}

$P_1, P_2, ., P_n \gets 1.0$ \Comment*[r]{Define learnable power values for each layer}

\For{$i \in [1, n]$}{ 
    Initializing optimizers, schedulers \\ 
    \For {each epoch}{
        \For {each data batch $x_{in} \sim P_X$}{
            $A_{out} = ||P_i \times (f_i \circ f_{i-1} \circ ... \circ f_1) (x_{in})||$
            
            $\mathcal{L}_{\phi}$ calculated according to section 2.1
            
            $f_i^*, P_i^* = \underset{f_i, P_i }{\mathrm{argmin}}(\mathcal{L}_{\phi})$\\
            
        }
    }
    
    \If{$i \neq 1$}{
    Initializing optimizers, schedulers \\ 
    \For {each epoch}{
        \For {each data batch $x_{in} \sim P_X$}{
            $A_{out} = ||P_i \times (f_i \circ f_{i-1} \circ ... \circ f_1) (x_{in})||$
            
            $\mathcal{L}_{\phi}$ calculated according to section 2.1
            
            $f_i^*,f_{i-1}^*,...,f_1^*,P_i^* = \underset{f_i,f_{i-1},..., the f_1, P_i }{\mathrm{argmin}}(\mathcal{L}_{\phi})$\\

        }
    }}
}

\end{algorithm}

\section{Comparison with Empirical Upper Bound}

\begin{figure}
    \centering
    \includegraphics[width=\textwidth]{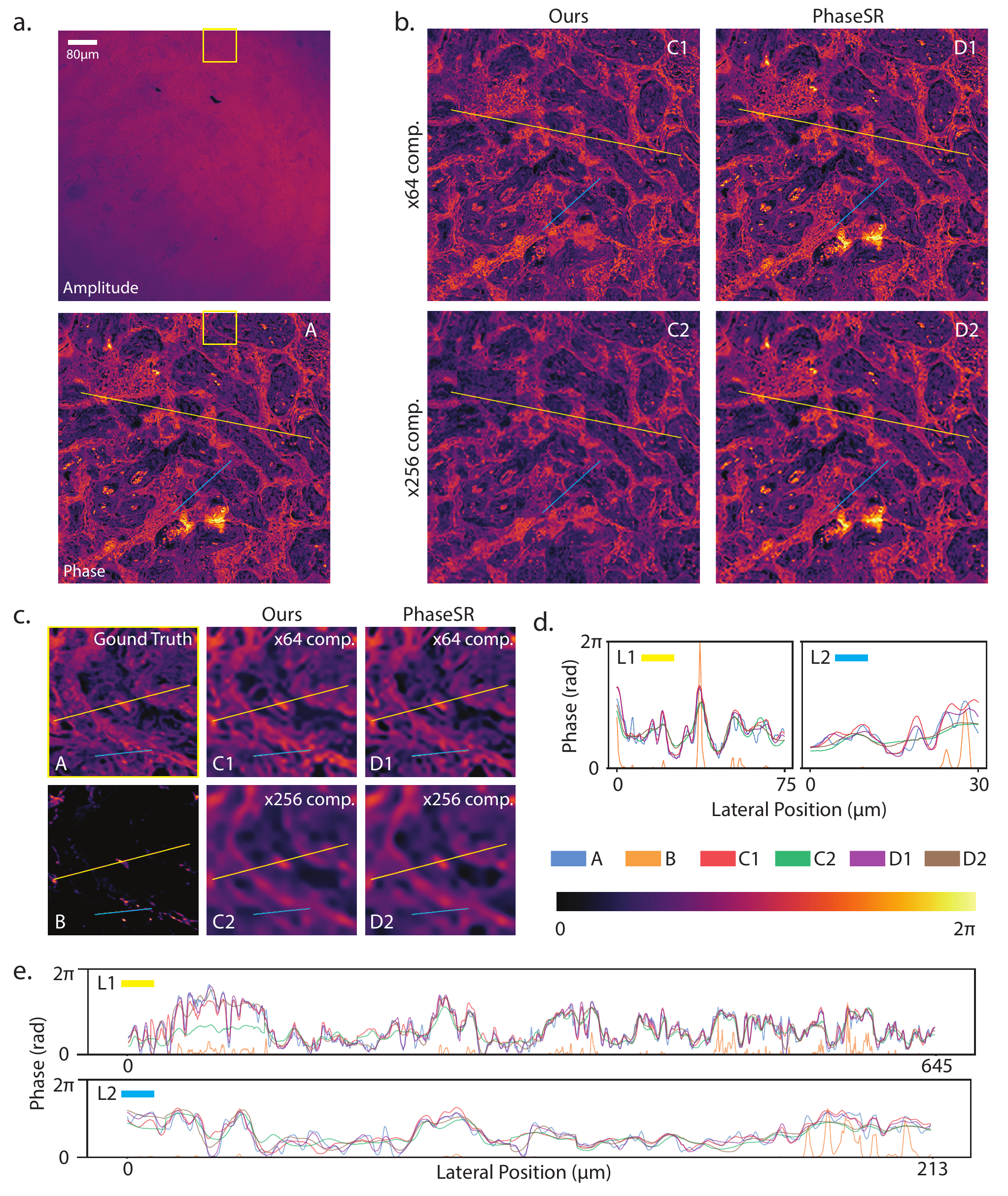}
    \caption{\textbf{Performance comparison on QPM tissue dataset}: a) A sample from test dataset. b) Reconstructions from our method and PhaseSR for $\times 64$ and $\times 256$ compressions. PhaseSR assumes a perfect all-optical phase to intensity transformation. The reconstruction network takes a ($\times 64$ or $\times 256$) downscaled ground truth phase as the input. Due to the perfect all-optical phase-to-intensity assumption, we treat PhaseSR as the empirical upper bound for our results. c) Magnified ground truth and corresponding results. Figure B shows the all-optical LFF reconstruction. d) Phase fluctuations along two lines on small field of views (C). e) Phase fluctuations along two lines on large field of views (B). Colors A-D2 are ground truth, all-optical LFF, ours with $\times 64$ compression, ours with $\times 256$ compression, PhaseSR with $\times 64$ compression, PhaseSR with $\times 256$ compression (please refer to table \ref{table:results_tissue}).}
    \label{fig:results_tissue}
\end{figure}

We further tested our method on a tissue dataset (section 5). All the hyper-parameters were similar to the HeLa experiments. We compared our method with a baseline we call \textit{PhaseSR}. PhaseSR is a hypothetical phase reconstruction method that assumes perfect phase-to-intensity transformation. This removes the bottleneck of converting phase to intensity optically. Therefore only bottleneck the end-to-end pipeline has to handle is reconstructing missing information due to the downsampling of the field. The input to the super-resolution network was a downscaled phase distribution where the phase values were distributed in $[0, 2\pi]$. We considered $\times 64$ and $\times 256$ downscaling and compared the qualitative (Fig. \ref{fig:results_tissue}) and quantitative (Table \ref{table:results_tissue}) results with our method. Due to the ideal phase-to-intensity transformation in PhaseSR, we considered it as the empirical upper bound for the results.

Poor performances in all-optical baselines indicate that phase retrieval of the tissue dataset is more challenging than that of the HeLa cell dataset. The proposed methods also exhibited lower performance on this dataset. Furthermore, as anticipated, PhaseSR demonstrated higher quantitative performance compared to the proposed method. However, the qualitative performance similarity between the proposed method and PhaseSR suggests that compression plays a more significant role in degrading performance than phase-to-intensity conversion.

\begin{table}[H]
\centering
\begin{tabular}{cclccccc}
\hline 
\multirow{2}{*}{\rotatebox{90}{Experiment}} & \multirow{2}{*}{\rotatebox{90}{Optical Net}} & \multirow{2}{*}{\rotatebox{90}{Compression}} & \multirow{2}{*}{\rotatebox{90}{Finetune}} & \multicolumn{2}{c}{\begin{tabular}[c]{@{}c@{}}\\ \\ \\ full FoV/ patch FoV \\ Metrics\end{tabular}} \\ [45pt] \cline{5-6} 
 &   &  &  &  \textbf{PSNR} & \textbf{SSIM} \\ \hline

B & \multicolumn{3}{l}{All optical LFF} & \begin{tabular}[c]{@{}c@{}}11.5446/\\ 12.3976\end{tabular} & \begin{tabular}[c]{@{}c@{}}0.0640/\\ 0.0632\end{tabular} \\  \hline

C1 & \multirow{4}{*}{LFF} & \multirow{2}{*}{64} & \cmark  & \begin{tabular}[c]{@{}c@{}}22.1981/\\ 24.2548\end{tabular} & \begin{tabular}[c]{@{}c@{}}0.7134/\\ 0.7147\end{tabular} \\ \cline{4-8}
C2 &  & \multirow{2}{*}{256} & \cmark  & \begin{tabular}[c]{@{}c@{}}19.5154/\\ 21.1397\end{tabular} & \begin{tabular}[c]{@{}c@{}}0.5555/\\ 0.5563\end{tabular} \\  \hline

D1 & \multirow{4}{*}{PhaseSR} & \multirow{2}{*}{64} & - &\begin{tabular}[c]{@{}c@{}}24.8808/\\ 27.6686\end{tabular} & {\begin{tabular}[c]{@{}c@{}}0.7759/\\ 0.7735\end{tabular}} \\ \cline{4-8} 
D2 &  & \multirow{2}{*}{256} & - & \begin{tabular}[c]{@{}c@{}}22.539/\\ 23.7128\end{tabular} & \begin{tabular}[c]{@{}c@{}}0.6043/\\ 0.6036\end{tabular} \\  \hline

\end{tabular}
\caption{\textbf{Performance comparison on tissue dataset}: We compare the performance of our method with PhaseSR, for $\times 64$ and $\times 256$ compressions. Since PhaseSR assumes a perfect all-optical phase-to-intensity transformation, we treat it as the empirical upper bound for our results. Refer to Fig. \ref{fig:results_tissue} for qualitative comparison} 

\label{table:results_tissue}
\end{table}

